\documentclass[preprint,aps]{revtex4}

\usepackage{graphicx}
\begin{document}

% The title goes here
%
%
\title{Unusual Fermi Surface Sheet-Dependent Band Splitting in Sr$_2$RuO$_4$ Revealed by High Resolution Angle-Resolved Photoemission}
%
%

%The list of authors
%
\author{Shanyu Liu$^{1}$, Hongming Weng$^{1}$, Daixiang Mou$^{1}$, Wentao Zhang$^{1}$, Quansheng Wu$^{1}$, Junfeng He$^{1}$, Guodong Liu$^{1}$, Lin Zhao$^{1}$, Haiyun Liu$^{1}$, Xiaowen Jia$^{1}$, Yingying Peng$^{1}$, Shaolong He$^{1}$, Xiaoli Dong$^{1}$, Jun Zhang$^{1}$, Z. Q. Mao$^{2}$,  Chuangtian Chen$^{3}$, Zuyan Xu$^{3}$, Xi Dai$^{1}$, Zhong Fang$^{1}$ and X. J. Zhou $^{1,*}$}

\affiliation{
\\$^{1}$National Laboratory for Superconductivity, Beijing National Laboratory for Condensed Matter Physics, Institute of Physics, Chinese Academy of Sciences, Beijing 100190, China
\\$^{2}$Department of Physics and Engineering Physics, Tulane University, New Orleans, Louisiana 70118, USA
\\$^{3}$Technical Institute of Physics and Chemistry, Chinese Academy of Sciences, Beijing 100190, China
}
\date{May 22, 2012}

\begin{abstract}

High resolution angle-resolved photoemission measurements have been carried out on Sr$_2$RuO$_4$. We observe clearly two sets of Fermi surface sheets near the ($\pi$,0)-(0,$\pi$) line which are most likely attributed to the surface and bulk Fermi surface splitting of the $\beta$ band. This is in strong contrast to the nearly null surface and bulk Fermi surface splitting of the $\alpha$ band although both have identical orbital components. Extensive band structure calculations are performed by considering various scenarios, including structural distortion, spin-orbit coupling and surface ferromagnetism. However, none of them can explain such a qualitative difference of the surface and bulk Fermi surface splitting between the $\alpha$ and $\beta$ sheets. This unusual behavior points to an unknown order on the surface of Sr$_2$RuO$_4$ that remains to be uncovered. Its revelation will be important for studying and utilizing novel quantum phenomena associated with the surface of Sr$_2$RuO$_4$ as a result of its being a possible p-wave chiral superconductor and a topological superconductor.

\end{abstract}


\maketitle

The Sr$_2$RuO$_4$ superconductor with a T$_c$$\sim$1.5 K\cite{MaenoNature} has attracted much attention since its first discovery in 1994 because of its normal Fermi liquid behaviors\cite{SRO_FL} and unconventional superconductivity\cite{SROReview,MaenoRev}. It is the only superconductor without copper but with a layered perovskite crystal structure similar to that of the copper-oxide high temperature superconductors\cite{MaenoNature}. Unconventional {\it p}-wave triplet superconductivity was proposed\cite{TMRice} and experimentally tested in various experiments\cite{Luke,KIshida,KDNelson,JXia}. The further identification of possible chiral p$_x$+ip$_y$ superconducting wave function\cite{Kidwingira,EdgeStateSRO} renders Sr$_2$RuO$_4$ classified as a time reversal symmetry breaking topological superconductor\cite{SCZhangReview,SROReview,MaenoRev}.  This becomes particularly interesting at present because of its close relation with Majorana Fermions and non-Abelian statistics which have a potential application for topological quantum computation\cite{DaSarma}.

The underlying electronic structure is essential to understand the unconventional superconductivity and other surface and interface-related quantum phenomena in Sr$_2$RuO$_4$. So far the electronic structure of Sr$_2$RuO$_4$ has been well investigated from both the band structure calculations\cite{LDATOguchi,LDASingh,IMazin1} and experimental measurements\cite{Mackenzie,Bergemann,DamascelliPRL,KShenSurface,HisorSRO}. The low-energy electronic structure of Sr$_2$RuO$_4$  is dictated by the Ru 4d orbitals\cite{LDATOguchi,LDASingh,IMazin1}. Among them, the d$_{xz}$ and d$_{yz}$ orbitals give rise to two sets of quasi-one-dimensional Fermi surface sheets (vertical dashed lines and horizontal dashed lines in Fig. 1a, respectively) which, after hybridization, lead to a hole-like Fermi surface sheet near X($\pi$,$\pi$) and an electron-like sheet near $\Gamma$(0,0) (Fig. 1a). The in-plane d$_{xy}$ orbital produces a two-dimensional electron-like Fermi surface sheet around $\Gamma$ (Fig. 1a). Quantum oscillation measurements\cite{Mackenzie,Bergemann}, as well as angle-resolved photoemission (ARPES) measurements\cite{DamascelliPRL}, are consistent with such band structure calculations\cite{LDATOguchi,LDASingh,IMazin1}.  It was later found that the Sr$_2$RuO$_4$ surface can get reconstructed with a rotation of the RuO$_6$ octahedra on the surface along the {\it c} axis\cite{Matzdorf}. Such a surface reconstruction gives rise to three surface-related Fermi surface sheets (blue lines in Fig. 1a). Furthermore, this RuO$_{6}$ rotation gives rise to $\sqrt{2}\times\sqrt{2}$ surface reconstruction\cite{Matzdorf} that produces folded shadow bands (green lines in Fig. 1a) corresponding to the surface Fermi surface sheets\cite{DamascelliPRL,KShenSurface}. Therefore, the three sets of Fermi surface sheets corresponding to a bulk, a surface and a folded surface, have been established as a general Fermi surface picture for Sr$_2$RuO$_4$.

%%%%2. Highlight of the present paper
In this paper, we report a Fermi surface picture that deviates from the above well-established case. By performing high resolution ARPES measurements on Sr$_2$RuO$_4$, we have identified two sets of Fermi surface sheets along the ($\pi$,0)-(0,$\pi$) line\cite{SYNote,BorisenkoSRO}. These can reasonably be attributed to the surface and band splitting of the $\beta$ band.  However, this is in strong contrast to the nearly zero surface and bulk Fermi surface splitting of the $\alpha$ sheet although they share identical orbital character. Such a disparate surface and bulk Fermi surface splitting between the $\alpha$ and $\beta$ bands can not be explained by the known scenarios, including structural distortion, spin-orbit coupling and surface ferromagnetism.  The unusual behavior points to the existence of some unknown order  to be identified on the surface of Sr$_2$RuO$_4$ that can break the equivalence of the $\alpha$ and $\beta$ bands.

%%%%3. Experimental description:  ARPES and Sr2RuO4 sample
High resolution angle-resolved photoemission measurements were carried out on our lab system equipped with a Scienta R4000 electron energy analyzer\cite{GDLiu}. We use helium discharge lamp as the light source which can provide photon energies of h$\nu$= 21.218 eV (Helium I).  The energy resolution was set at  4$\sim$10 meV  and the angular resolution is $\sim$0.3 degree. The Fermi level is referenced by measuring on a clean polycrystalline gold that is electrically connected to the sample. The Sr$_2$RuO$_4$ crystals were grown by the traveling solvent floating zone method\cite{ZQMaoGrowth}.  The crystal was cleaved {\it in situ} and measured in vacuum with a base pressure better than 5$\times$10$^{-11}$ Torr.  The results reported are reproducible from many separate measurements.

%%%%1. Observation of extra Fermi surface sheets
The high resolution ARPES measurement of Sr$_2$RuO$_4$ (Fig. 1b) reveals new features that are not observed in the previous ARPES measurements\cite{DamascelliPRL,KShenSurface,HisorSRO}. Fig. 1b shows the Fermi surface mapping of Sr$_2$RuO$_4$ cleaved at 20 K and measured at 20 K. The band structure along several typical cuts are presented in Fig. 2. Considering both Fig. 1b and Fig. 2, we obtain new and experimentally determined Fermi surface topology of Sr$_2$RuO$_4$ shown in Fig. 1c.

Overall speaking, compared with the well-accepted Fermi surface topology in Fig. 1a, in spite of an overall consistence with the previous ARPES measurements and LDA calculations, there are a couple of obvious differences.   (1). Only one Fermi surface sheet is resolved for the $\alpha$ band (Fig. 1c) which is in contrast to the two Fermi surface sheets expected due to bulk and surface splitting (Fig. 1a). In fact, upon a close examination on the band structure, it is clear that the $\alpha$ Fermi surface sheet is actually composed of two bands with nearly the same Fermi momentum k$_F$ (Fig. 2c). In fact, over the entire Fermi surface sheet, the bulk and surface Fermi surface  sheets for the $\alpha$ band overlap (Fig. 2c). (2). Compared with the previous ARPES measurements (Fig. 1a), the most obvious difference is the observation of four Fermi surface sheets\cite{SYNote,BorisenkoSRO} forming two pairs (3 and 6, and 4 and 5) along the ($\pi$,0)-(0,$\pi$) line with the sheets 6 and 5 being the corresponding shadow bands of the sheets 3 and 4, respectively. In the previous ARPES measurements\cite{DamascelliPRL,KShenSurface,HisorSRO}, usually only one pair of such Fermi surface sheets was clearly observed. Compared with the expected Fermi surface where there are three shadow bands (Fig. 1a),  the observation of four shadow bands(Fig. 1b and 1c) indicates there is one additional band that is resolved.

%%%%2. The behaviors of these extra bands: Temperature dependence and aging effect
%%%%6: Fig. 3: MDCs for fresh and aged surface
%%Make SECOND argument: Which one is splitted, from where?

To gain further insight on the nature of these two sets of bands, we carried out temperature dependence and aging experiment on Sr$_2$RuO$_4$. It is well-known that the Sr$_2$RuO$_4$ surface can get aged with time and the aging effect can get enhanced at high temperature. After aging, the surface bands can be strongly suppressed and the bulk Fermi surface may become dominant\cite{DamascelliPRL,HisorSRO,SYLiu}. Fig. 3 shows the Fermi surface of Sr$_2$RuO$_2$ cleaved at 20 K and measured then at different temperatures, first warming up gradually to 100 K [Figs. 3a-3e)] and then cooled down to 20 K again (Fig. 3f).  Fig. 3h and Fig. 3i shows the band structure initially measured at 20 K and finally measured at 20 K after aging, respectively, along the momentum cut shown in Fig. 3a; the corresponding momentum distribution curves (MDCs) at E$_F$ are shown in Fig. 3g for different temperatures . It is clear from Fig. 3 that:  (1). The existence of both sets of bands is visible up to at $\sim$80 K. Note that during the warming process the suppression of the bands 4 and 5 is due to both the temperature effect and aging effect.  (2). With increasing temperature and aging, the initially strong band 4 gets weaker and eventually almost disappears, while the initially weak band 3 stays at high temperature and after aging.  This indicates that the band 4 is predominantly of surface nature while the band 3 possesses obvious bulk nature. (3). After aging, the initial band 1 decreases in intensity and shifts its position significantly after aging, but the band 2 keeps its position after aging even though its intensity also decreases.

It is straightforward to assign the band 1 and 1$^{'}$ as due to the bulk and surface band of the $\alpha$ band, and the band 2 and 2$^{'}$ as due to the bulk and surface bands of the $\gamma$ band. The main issue to be addressed is the nature of the Fermi surface  sheets 3 and 4.
There are two obvious possibilities emerged.  On the one hand, if we believe bulk bands do not produce shadow bands, as normally considered, then both bands 3 and 4 can only be surface bands.  On the other hand, if we believe the bulk band can also produce shadow band, then one may attribute bands 3 and 4 as the bulk and surface $\beta$ band, respectively. In light of the surface nature of the band 4 and bulk nature of the band 3 as demonstrated in Fig. 3, and considering the $\beta$ bulk and surface band splitting in Fig. 1a, it appears that the second possibility is more plausible.  However, this assignment leaves a couple of inherent inconsistencies. (1). There is an obvious difference on the location of the Fermi surface sheets 3 and 4; they are obviously much closer to the ($\pi$,0)-(0,$\pi$) line than those in Fig. 1a;  (2). There is a dramatic difference of the bulk and surface band splitting between the $\alpha$ and $\beta$ bands. While the $\alpha$ band shows nearly zero bulk and surface Fermi surface splitting, the splitting for the $\beta$ band is obvious and significant.  As discussed above, because the $\alpha$ band and $\beta$ band originate from the same orbitals of d$_{xz}$ and d$_{yz}$, one would expect they exhibit similar behaviors, as expected to split similarly from the band structure calculations (Fig. 1a).

%%While the first two issues seem to be on a quantitative level that are in principle still possible, the third issue is significant on the
%%qualitative basis.

In order to further understand the origin of the bands 3 and 4, and the associated puzzle of Fermi surface-dependent band splitting, we have carried out comprehensive band structure calculations by considering various scenarios.  We choose repeated slabs consisting of three-perovskite-unit layers  separated by vacuum region (Fig. 4a) to take into account both the bulk and surface electronic structures; such a thin slab calculation is proven to be reliable in simulating the surface\cite{Matzdorf}. The electronic structure represented by the middle layer is identical to the bulk electronic structure in the d-band region because of weak d-d hopping across the SrO insulating layer\cite{Matzdorf}.  In the simplest case where the three layers in the slab all assume the Sr$_2$RuO$_4$ bulk structure, three sets of Fermi surface sheets can be generated ($\alpha$, $\beta$ and $\gamma$). Each set is composed of one bulk and two surface Fermi surface sheets; the two surface sheets exhibit slight splitting due to coupling between the top and bottom surfaces\cite{SYLiu}.

The band structure shows a dramatic change with the rotation of the RuO$_6$ octahedra on the surface introduced (Fig. 4b). The bulk and surface Fermi surface sheets show obvious splitting for all three bands. Particularly, the $\gamma$ band splits into one hole-like bulk band and an electron-like surface band\cite{KShenSurface}.  We took a rotation angle of 6 degrees which is close to the one determined\cite{Matzdorf} and the calculated results (Fig. 4c) are similar to that obtained before (Fig. 1a)\cite{KShenSurface}. It is clear that the bulk and surface $\alpha$ band show an obvious splitting, and its splitting is similar to that of the $\beta$ bulk and surface Fermi surface sheets.  Variation of the octahedra rotation angle does not produce a different splitting of the surface and bulk bands between the $\alpha$ and $\beta$ sheets.  Therefore, the RuO$_6$ octahedra rotation can not explain different bulk and surface Fermi surface splitting between the $\alpha$ and $\beta$ sheets. Moreover, the location of measured bands 3 and 4 deviates considerably from the calculations: they are much closer to the ($\pi$,0)-(0,$\pi$) line than the calculated ones.

%%In particular, the range of the rotation angle is nearly fixed due to the experimental measurement\cite{Matzdorf} and the band structure
%%calculations.

Since Sr$_2$RuO$_4$ involves spin-orbit coupling\cite{Haverkort}, it is natural to examine whether this unusual Fermi surface-dependent band splitting could be due to spin-orbit coupling, as suggested before\cite{BorisenkoSRO}. Fig. 4d shows the calculated Fermi surface of Sr$_2$RuO$_4$ by considering the spin-orbit coupling\cite{Haverkort}. It is clear that the spin-orbit coupling causes only a subtle change of the Fermi surface. The calculated $\beta$ Fermi surface sheets shift slightly toward ($\pi$,0)-(0,$\pi$) line. However, on a large scale, it does not solve the two issues concerning the $\beta$ band location and the unusual Fermi surface-dependent band splitting.

Because the bandwidth narrowing of the d$_{xy}$ orbital at the Sr$_2$RuO$_4$ surface due to RuO$_6$ octahedral rotation increases the density of states at the Fermi level, it was suggested that ferromagnetic ground state may be stabilized at the surface\cite{Matzdorf}.  Such a surface magnetism is expected to cause further splitting of the initial surface and bulk bands, a scenario that were examined previously but without revealing clear evidence\cite{KShenSurface}. The revelation of extra bands in the present high resolution ARPES measurements prompted us to explore this scenario again in more detail. Fig. 4e shows the calculated Fermi surface involving surface ferromagnetism. In this case, the surface magnetic moment is calculated to be 1.0 $\mu_{B}$$\slash$Ru which is strongly enhanced when compared to the 0.4 $\mu_{B}$$\slash$Ru calculated for the bulk\cite{Matzdorf}. Each of the initial surface and bulk Fermi surface sheets is further split into two bands with one spin-up and the other spin-down. The splitting also causes the band energy position shift which moves some bands above or below the Fermi level so the final Fermi surface sheets are not simply twice that of the original number of bands. Altogether, the surface magnetism causes drastic band structure and Fermi surface change, both in the number of the Fermi surface sheets and the shifting of the Fermi surface location (Fig. 4e). As seen in Figs. 1b and 1c, such dramatic alteration of Fermi surface is not consistent with the observed Fermi surface.

It is known that substitution of Sr with Ca in Sr$_2$RuO$_4$ can cause the tilting of the RuO$_6$ octahedra in addition to its rotation along the c-axis\cite{OFriedt,ZFang}. Although there is no direct evidence of RuO$_6$ tilting revealed so far in Sr$_2$Ru$O_4$, its presence in its neighbor (Sr$_{2-x}$Ca$_{x}$)RuO$_4$ indicates there may be such a tilting tendency. We therefore investigated how the electronic structure is modified if such an octahedra tilting occurs in Sr$_2$RuO$_4$. As seen in Fig. 4f, with an introduction of a slight octahedra tilting (3 degree), the bulk Fermi surface shows little change, but the surface Fermi surface exhibits a dramatic change. In particular, the portion of Fermi surface near ($\pi$,0) and (0,$\pi$) regions disappears for both the $\beta$ surface Fermi surface and $\gamma$ surface Fermi surface, forming a closed Fermi surface near the middle of the Brillouin quadrant. In addition, some additional Fermi surface appears near (0,$\pi$) along the (0,0)-(0,$\pi$) line but not along the (0,0)-($\pi$,0) line. As compared with the measured Fermi surface (Figs. 4f), such an octahedra tilting can not account for the location of the Fermi surface sheets 3 and 4, nor can it explain the difference in the band splitting of $\alpha$ and $\beta$ bands.

It is clear that, in order to understand why the $\beta$ band shows a disparate splitting from that of the $\alpha$ band, it requires a mechanism that can break the equivalency between the $\alpha$ and $\beta$ Fermi surface sheets.  The scenarios we have discussed above, including the structural distortion (octahedral rotation and tilting), spin-orbit coupling, and possible surface ferromagnetism can not explain such an unusual behavior. Since Sr$_2$RuO$_4$ is known to have strong electron correlation\cite{SROReview} as evidenced by appreciable band renormalization\cite{HisorSRO}, it is intriguing to investigate the effect of electron correlation on the band structure of Sr$_2$RuO$_4$.  While this inclusion may quantitatively modify the results, it remains unclear how it can break the equivalence between the $\alpha$ and $\beta$ Fermi surface sheets.  As the electronic inhomogeneity can be present in the correlated electron systems such as stripes, it is also interesting to see whether a stripe order can alter the equivalence\cite{SRaghu}. While the possible formation of one-dimensional structure may break the equivalency between the d$_{xz}$ and d$_{yz}$ orbitals that leads to a difference on the horizontal and vertical sections for a given $\alpha$ or $\beta$ Fermi surface sheet, we do not see its possibility to change the equivalence between the $\alpha$ and $\beta$ Fermi surface sheets. We have also considered the possibility of surface charging by some extra carrier doping, for example, by formation of some surface defects. However, after a careful examination, we found that such a surface charging will not break the equivalence of the surface and bulk Fermi surface splitting of the $\alpha$ and $\beta$ bands either.

%%%%8.Summary

In summary, even after we have exhausted all possible scenarios known to the best of our knowledge, conventional or exotic, we can not resolve the obvious discrepancies between the experiment and theoretical calculations.  This indicates that there may be some order hidden on the surface of Sr$_2$RuO$_4$. Our present work has unveiled the existence of such an unknown order, however, its exact nature remains to be further explored.  Such an order is related to the surface of Sr$_2$RuO$_4$ which can break the equivalence between the $\alpha$ and $\beta$ bands.  As it has been shown that Sr$_2$RuO$_4$ is a possible topological superconductor which involves interesting edge state on the surface, it is important to understand the surface electronic structure beforehand in order to reveal and utilize many interesting quantum phenomena near the surface and interface.  We hope the present work will stimulate further effort and ideas to reveal the order that can account for the unusual phenomena we have observed.

%%It is intriguing to note that, after so many years of extensive and deep investigations on Sr$_2$RuO$_4$, its electronic structure remains to
%%pose challenges to our understanding.

XJZ thanks the funding support from the NSFC (Grant No. 10734120) and the MOST of China (973 program No: 2011CB921703).

$^{*}$Corresponding author (XJZhou@aphy.iphy.ac.cn)

\newpage

\begin{figure*}[tbp]
%%\begin{figure}[tbp]
\begin{center}
\includegraphics[width=1.00\columnwidth,angle=0]{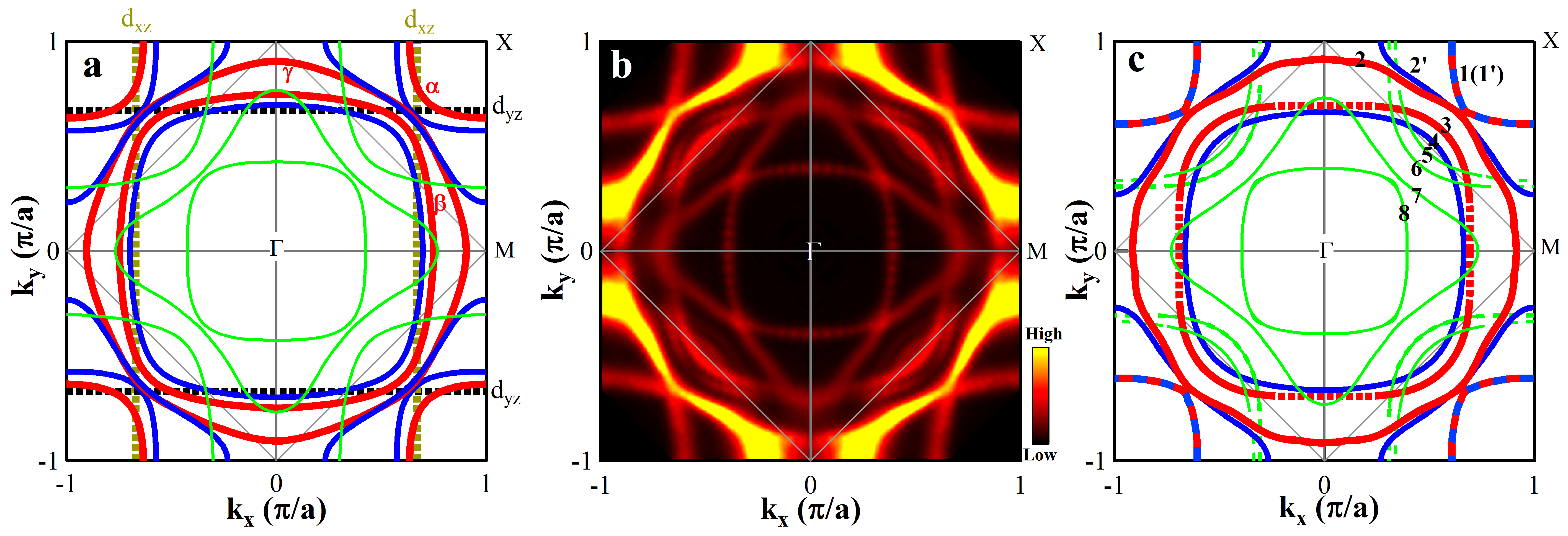}
\end{center}
\caption{Fermi surface of Sr$_2$RuO$_4$.  (a). LDA calculated Fermi surface of Sr$_2$RuO$_4$\cite{KShenSurface} with bulk Fermi surface sheets $\alpha$, $\beta$ and $\gamma$ (thick red lines), corresponding surface Fermi surface sheets (blue lines) and the umklapp Fermi surface sheets of the surface Fermi surface (green lines). (b). Measured Fermi surface mapping of Sr$_2$RuO$_4$ cleaved at 20 K and measured at 20K. (c). Fermi surface deducted from Fig. 1b. The observed Fermi surface sheets are labeled by numbers.
}
%%\end{figure}
\end{figure*}

%%\begin{figure*}[tbp]
\begin{figure}[b]
\begin{center}
\includegraphics[width=1.00\columnwidth,angle=0]{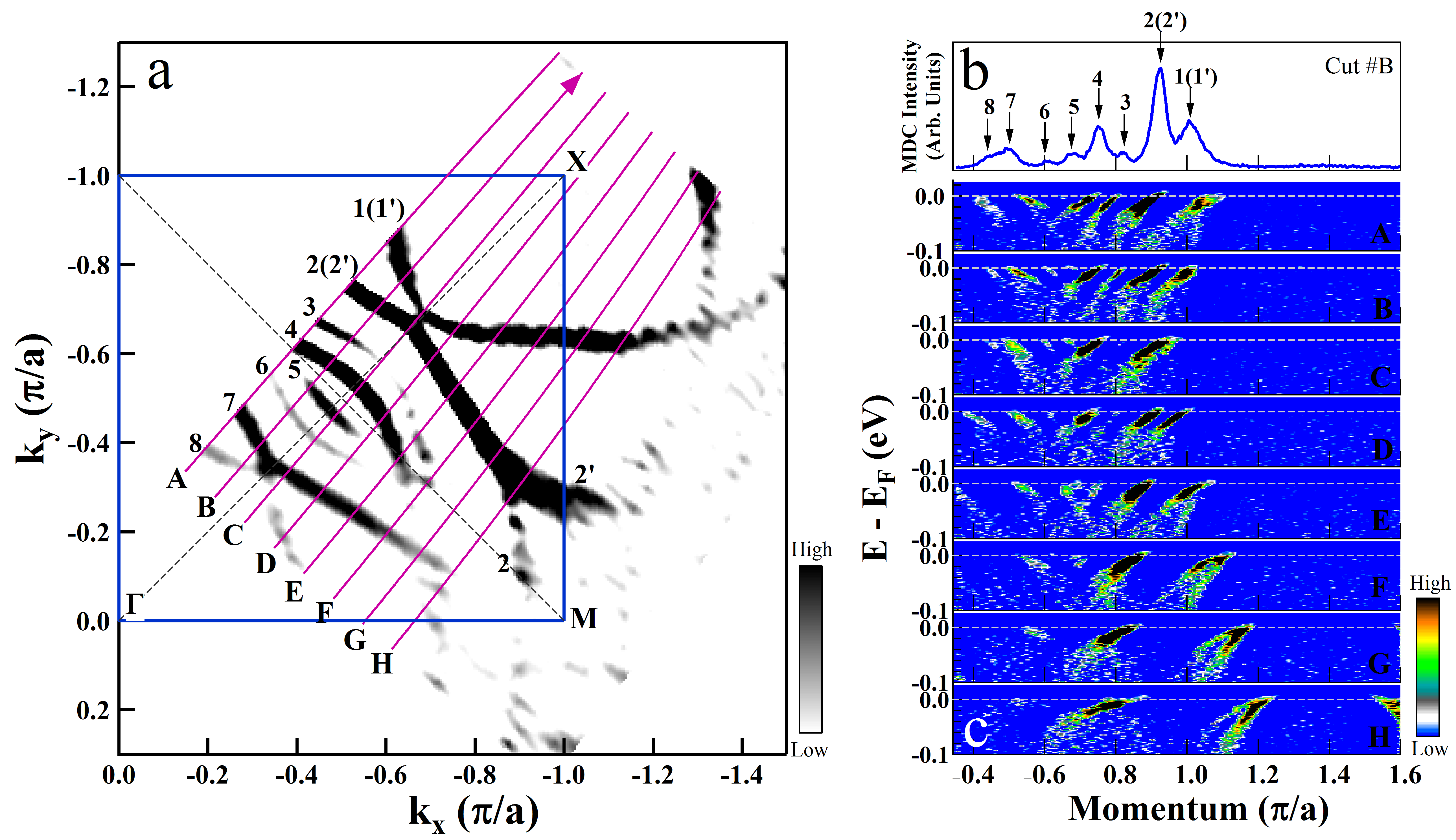}
\end{center}
\caption{Band structure of Sr$_2$RuO$_4$ cleaved at 20 K and measured at 20 K. (a). Fermi surface mapping and the location of several typical momentum cuts. The intensity map is obtained by first performing second derivative of the original momentum distribution curves before integrating over a small energy windows [-3meV,3meV] with respect to the Fermi level (E$_F$). The corresponding band structure for the momentum cuts A to H are shown in (c) which are the MDC second derivatives of the original photoemission data.  The corresponding momentum distribution curve for the Cut$\#$B at the Fermi level is shown in (b) with the observed peaks labeled.
}
\end{figure}
%%\end{figure*}

\begin{figure*}[tbp]
%%\begin{figure}[t]
\begin{center}
\includegraphics[width=1.00\columnwidth,angle=0]{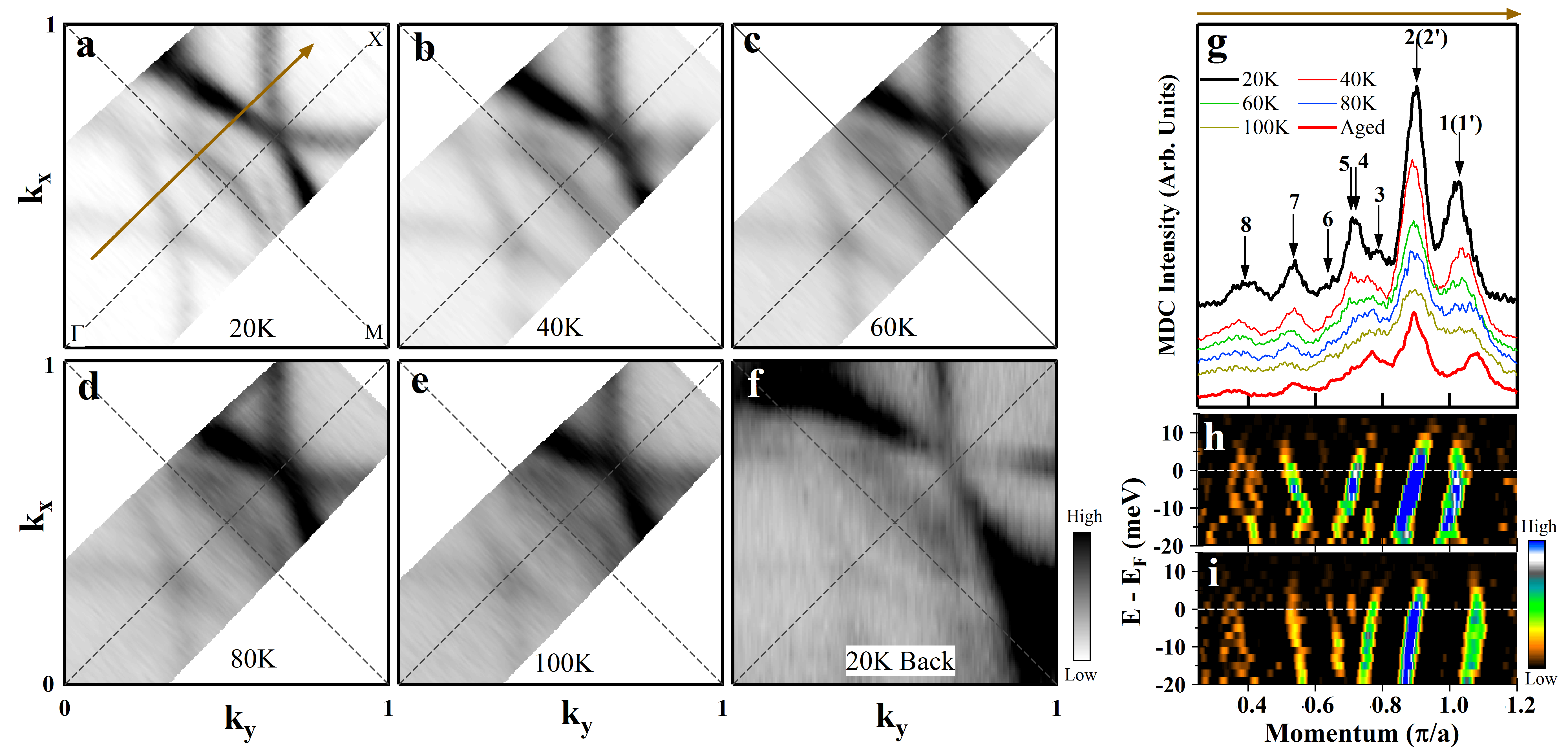}
\end{center}
\caption{Evolution of Fermi surface and band structure with temperature and time.  (a-e) show Fermi surface mapping of Sr$_2$RuO$_4$ cleaved at 20 K and measured at different temperatures, going from 20 K(a), 40 K(b), 60 K(c), 80 K(d), to 100 K(e).  Then the sample was cooled back to 20 K and the Fermi surface was measured again on the aged surface as shown in (f). (e) shows MDCs at the Fermi level for the sample measured at different temperatures and for the aged sample. The location of the corresponding momentum cut is shown in Fig. 3a. The corresponding band structure for the fresh surface (h) and aged surface (i) are measured at 20 K. The images are MDC second derivative of the original data.
}
%%\end{figure}
\end{figure*}

\begin{figure*}[tbp]
%%\begin{figure}[b]
\begin{center}
\includegraphics[width=1.00\columnwidth,angle=0]{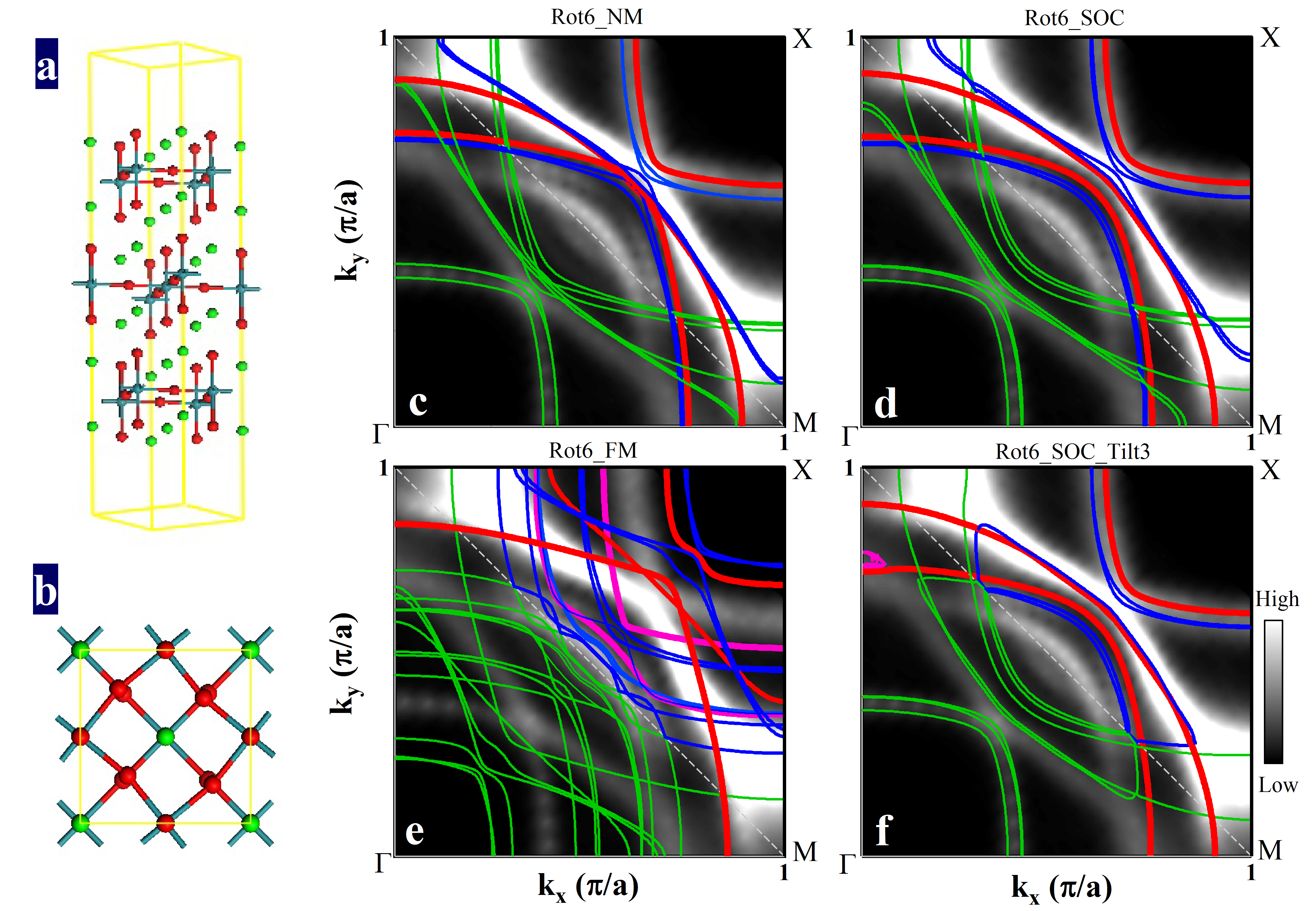}
\end{center}
\caption{Comparison between the measured and  calculated Fermi surface of Sr$_2$RuO$_4$. The calculation considers both the bulk Fermi surface and surface Fermi surface. (a) shows schematic three-slab of Sr$_2$RuO$_4$ unit.  (b) shows the crystal structure viewed from the c-axis which shows a rotation of the RuO$_6$ octahedron with an angle of $\theta$.  (c). Fermi surface of Sr$_2$RuO$_4$ with an octahedral rotation of 6 degree and without magnetic surface;   (d). Non-magnetic Fermi surface with octahedral rotation of 6 degree plus spin-orbit coupling; (e). Fermi surface of Sr$_2$RuO$_4$ with an octahedral rotation of 6 degree and an assumed ferromagnetic surface;  (f). Fermi surface of Sr$_2$RuO$_4$ with an octahedral rotation of 6 degree, without magnetic surface, and with an assumed octahedral tilt of 3 degree.
}
%%\end{figure}
\end{figure*}

\end{document}